\documentclass[aps,prl,showpacs,superscriptaddress,nofootinbib,twocolumn,10pt]{revtex4}
\usepackage{color,amsmath,amssymb,graphicx,latexsym,lineno,amsfonts}
\usepackage{threeparttable,multirow,txfonts,subfigure,booktabs}

\graphicspath{{fig/}}

\begin{document}

\title{Diverse properties of electron Forbush decreases revealed by the Dark Matter Particle Explorer}

\author{F.~Alemanno}
\affiliation{Dipartimento di Matematica e Fisica E. De Giorgi, Universit\`a del Salento, I-73100, Lecce, Italy}
\affiliation{Istituto Nazionale di Fisica Nucleare (INFN) - Sezione di Lecce, I-73100, Lecce, Italy}

\author{Q.~An}
\affiliation{State Key Laboratory of Particle Detection and Electronics, University of Science and Technology of China, Hefei 230026, China}
\affiliation{Department of Modern Physics, University of Science and Technology of China, Hefei 230026, China}

\author{P.~Azzarello}
\affiliation{Department of Nuclear and Particle Physics, University of Geneva, CH-1211, Switzerland}

\author{F.~C.~T.~Barbato}
\affiliation{Gran Sasso Science Institute (GSSI), Via Iacobucci 2, I-67100 L’Aquila, Italy}
\affiliation{Istituto Nazionale di Fisica Nucleare (INFN) - Laboratori Nazionali del Gran Sasso, I-67100 Assergi, L’Aquila, Italy}

\author{P.~Bernardini}
\affiliation{Dipartimento di Matematica e Fisica E. De Giorgi, Universit\`a del Salento, I-73100, Lecce, Italy}
\affiliation{Istituto Nazionale di Fisica Nucleare (INFN) - Sezione di Lecce, I-73100, Lecce, Italy}

\author{X.~J.~Bi}
\affiliation{University of Chinese Academy of Sciences, Beijing 100049, China}
\affiliation{Particle Astrophysics Division, Institute of High Energy Physics, Chinese Academy of Sciences, Beijing 100049, China}

\author{H.~Boutin}
\affiliation{Department of Nuclear and Particle Physics, University of Geneva, CH-1211, Switzerland}

\author{I.~Cagnoli}
\affiliation{Gran Sasso Science Institute (GSSI), Via Iacobucci 2, I-67100 L’Aquila, Italy}
\affiliation{Istituto Nazionale di Fisica Nucleare (INFN) - Laboratori Nazionali del Gran Sasso, I-67100 Assergi, L’Aquila, Italy}

\author{M.~S.~Cai}
\affiliation{Key Laboratory of Dark Matter and Space Astronomy, Purple Mountain Observatory, Chinese Academy of Sciences, Nanjing 210023, China}
\affiliation{School of Astronomy and Space Science, University of Science and Technology of China, Hefei 230026, China}

\author{E.~Casilli}
\affiliation{Gran Sasso Science Institute (GSSI), Via Iacobucci 2, I-67100 L’Aquila, Italy}
\affiliation{Istituto Nazionale di Fisica Nucleare (INFN) - Laboratori Nazionali del Gran Sasso, I-67100 Assergi, L’Aquila, Italy}

% ...

\collaboration{DAMPE Collaboration}\email{dampe@pmo.ac.cn}

\author{X.~Luo}
\affiliation{Shandong Institute of Advanced Technology, Jinan 250100, China}

\begin{abstract}
The Forbush decrease (FD) of cosmic rays is an important probe of the 
interplanetary environment disturbed by solar activities. In this work, 
we study the properties of 8 FDs electrons (including positrons) between 
2 GeV and 20 GeV from January, 2016 to March, 2024, with the Dark Matter 
Particle Explorer. The maximum decrease amplitudes of these events are 
about $30\%-15\%$, and the amplitudes reduce with energy. The recovery 
time of these events shows diverse behaviors of their energy-dependence. 
Some of them show strong energy-dependence, while some have a nearly 
constant recovery time. It has been shown that such diverse behaviors 
could be related with the geometry of the disturbed regions of the 
interplanetary space by coronal mass ejections (CME), represented by the 
combined effect of the CME velocity, angular spread, and ejection direction.
\end{abstract}


\maketitle

{\it Introduction.} ---
Galactic cosmic rays (GCRs) are affected by heliospheric plasma and the 
associated magnetic fields when propagating through the solar system, 
resulting in flux variations at different time scales (e.g., 
\cite{2000SSRv...93...55C,2013LRSP...10....3P}). 
One type of short-time variations, known as Forbush decreases (FDs) 
\cite{1937PhRv...51.1108F,1937Natur.140..316H}, is the quick drop and 
relatively slow recovery of the GCR fluxes over timescales of days to 
weeks. FDs are caused by energetic solar activities such as coronal mass 
ejections (CMEs) \cite{2010SoPh..263..223Y, 2000SSRv...93...55C} or 
corotating interaction regions (CIRs) \cite{1999AdSpR..23..567H}, 
which disturb the interplanetary space and block part of GCRs from 
reaching the detector. Fast CMEs can drive strong interplanetary shocks 
that also play an important role in generating FDs, such as the two-step 
FD including an initial decrease by the shock and a further reduction 
when the CME itself arrives \cite{2011JGRA..11611103J}. FDs have also 
been observed far away from the Earth 
\cite{2018A&A...611A..79G,1993JGR....98....1B,2019ApJ...874..167A}, 
indicating that they are universal phenomenon in the solar system.

FDs are widely studied by ground-based neutron monitors (NMs) or muon 
detectors (MDs), which measure secondary particles produced by interactions 
between hadronic GCRs and the atmosphere. The NM or MD data show the 
integral variations of the GCR intensities above specific energies, 
which depend on locations of the detectors due to a latitude dependent 
geomagnetic vertical rigidity cutoff (VRC; \cite{2005AdSpR..36.2012S}).
Nonetheless, direct measurements by space experiments, in spite that 
they are limited by effective areas, provide better composition and 
energy resolution, and are thus crucial to understanding the physics 
of FDs. Very few direct measurements of FDs were available, by balloon 
experiments and space satellites \cite{1961JGR....66.3950M,2018ApJ...853...76M,2021ApJ...920L..43A,2023SoPh..298....9L}.
The PAMELA experiment studied FDs of protons, helium nuclei, and 
electrons, showing charge-sign dependences of their recovery properties 
\cite{2018ApJ...853...76M}. Further investigation of a sample of proton 
FDs using PAMELA data showed that there are two types of recovery times, 
type I with significant energy dependence and type II without energy 
dependence \cite{2023SoPh..298....9L}. Similarly, ground-based detectors 
have identified two distinct types of recoveries
\cite{2007AdSpR..40..342J,2008JGRA..113.7102U,2016ApJ...827...13Z},
whose interpretation remains unclear. Using the daily proton fluxes 
measured by AMS-02 \cite{2021PhRvL.127A1102A}, Wang et al. studied the 
properties of FDs caused by CMEs and CIRs, finding that they do not differ 
much from each other \cite{2023ApJ...950...23W} .

Observations of FDs of the leptonic component are even rarer. Daily fluxes 
of electrons and positrons from May 2011 to November 2021 were published by 
AMS-02 \cite{2023PhRvL.130p1001A,2023PhRvL.131o1002A}, but no detailed 
analysis of the FDs has been carried out. An electron FD was reported by 
PAMELA \cite{2018ApJ...853...76M}, shows a faster recovery than protons and 
helium nuclei below 2 GV. The properties of an FD of cosmic ray electrons and 
positrons (hereafter CREs) were measured by the Dark Matter Particle Explorer 
(DAMPE) with significantly improved statistics and wide energy coverage, 
showing clearly energy-dependent recovery times between 2 and 20 GeV 
\cite{2021ApJ...920L..43A}. Note that the DAMPE detector itself does not
distinguish electrons from positrons due to the lack of magnetic field, and in
this work we add electrons and positrons together for the study. The fraction
of positrons in the sample is a few percents in our interested energy range 
\cite{2013PhRvL.110n1102A}. The decrease and recovery behavior observed by 
DAMPE for this event can be properly accounted for using a diffusion
barrier model \cite{2018ApJ...860..160L,2021ApJ...920L..43A}, the more clear 
underlying physics of FDs needs further exploration using more events. It is 
an intriguing question whether the recovery times of CRE FDs also experience 
two types of energy dependence as thoses of GCR nuclei.

In this work, we analyze the DAMPE CRE data from January 1, 2016 to 
March 31, 2024 to study the population of FDs. A total of eight large 
FDs are identified, each  with maximum decrease amplitudes exceeding $30\%$ of 
the average fluxes. The decrease and recovery properties of these events will 
be investigated in detail.

{\it DAMPE detector.} --- 
The DAMPE is a space detector for the observations of cosmic ray nuclei, 
electrons and positrons, and $\gamma$-ray photons 
\cite{2017APh....95....6C,2025arXiv251105409D}. 
The DAMPE detector is composed of four sub-detectors: a plastic scintillator 
detector (PSD; \cite{2017APh....94....1Y}) on the top for charge measurement, 
a silicon-tungsten tracker-converter (STK; \cite{2016NIMPA.831..378A}) for 
trajectory and charge measurement, a Bismuth-Germanium-Oxide imaging 
calorimeter (BGO; \cite{2012ChPhC..36...71Z}) for energy measurement and 
electron-proton separation, and a neutron detector 
(NUD; \cite{2020RAA....20..153H}) for auxiliary electron-proton separation. 
Serving as the key subdetector, the BGO calorimeter consists of 14 layers, 
each containing 22 bars for a total depth of about $32X_0$ which enables 
$\sim1.5\%$ energy resolution for $\sim10$ GeV CREs \cite{2017APh....95....6C}. 
Different from observing the corona graphs, DAMPE is very suitable for 
measurements of CREs with a relatively large acceptance of $\sim0.3$ m$^2$~sr 
and a high energy resolution. The detector was integrated in a dedicated 
satellite platform, and was launched into a sun-synchronous orbit at an 
altitude of 500 km on December 17, 2015 from the Jiuquan base 
\cite{2017APh....95....6C}.

\begin{figure*}
\centering
\includegraphics[width=1.0\textwidth]{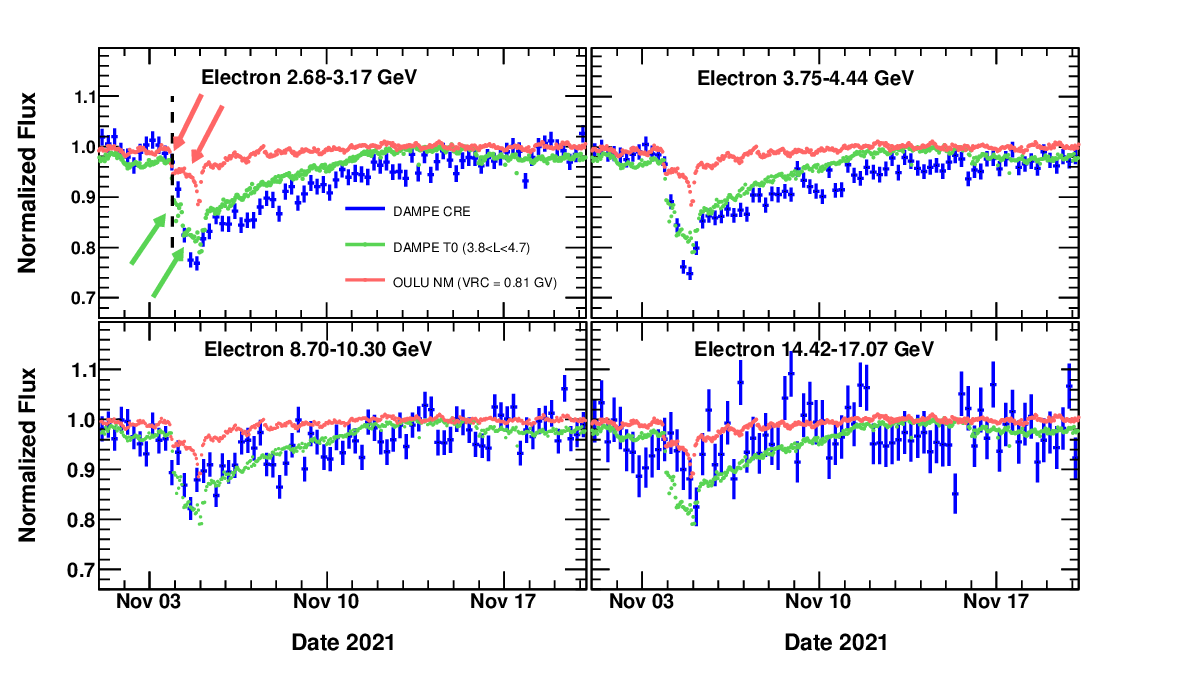}
\caption{Time profiles of CRE fluxes in four energy bins measured by DAMPE 
in November, 2021 (blue dots), compared with the DAMPE $T_0$ rate (green 
curves) and OULU NM data (red curves). The errorbars of the DAMPE CRE fluxes 
include both the statistical uncertainties and systematic uncertainties, 
added in quadrature. The 4 panels correspond to 4 different CRE energy bins 
as labelled. Arrows in the top-left panel show the two step decreases 
revealed by the NM data and DAMPE $T_0$ data, and the black dashed line 
shows the arrival time of the interplanetary shock.}
\label{fig:2021FD}
\end{figure*}

{\it CRE selection and flux calculation.} --- 
The housekeeping system of DAMPE records incident counts of events every 4 
seconds, referred to as $T_0$ counts, to monitor the condition of the data 
acquisition system \cite{2019RAA....19..123Z}. The $T_0$ trigger rate is a 
direct measure of the low-energy cosmic ray intensity, which varies 
from dozens Hz near the equator to tens of thousands Hz at the poles. 
Simulations indicate that $T_0$ triggering occurs when the energy of a particle
is higher than 100 MeV. FD candidates are selected using the DAMPE $T_0$ rate 
and the OULU NM count rate, both exhibit a rapid decrease plus a relatively 
slow recovery. Specifically, if the DAMPE $T_0$ rate decreases by more than 
10\% and the OULU NM count rate decreases over 3\% within 3 days, the 
corresponding time window is labelled as a possible FD event occurs. 
Then a cross match with the SOHO LASCO CME 
catalog\footnote{https://spase-metadata.org/NASA/Catalog/SOHO/LASCO/CME\_Catalog.html} is further required.
In total we find 8 candidate FDs according to this procedure. The basic 
information of these 8 FD events can be found in Table~S1 of the 
{\tt Supplemental Material}. For each candidate event, we select data from 
10 days before to 30 days after the starting time for the follow-up analysis.
CRE candidates are selected following the method described in 
Ref.~\cite{2021ApJ...920L..43A}. 
Data recorded during the passage of the South Atlantic Anomaly (SAA) region 
are excluded. Both events passing the high-energy trigger (HET) and low-energy
trigger (LET) \cite{2019RAA....19..123Z} conditions are used. 
The reconstructed particle charge \cite{2019APh...105...31D} is set 
to be $Z\le 1.7$, which rejects $\sim99\%$ of nuclei heavier than protons. 
We further require that the shower maximum is not at the edge of the first 
six BGO layers, and the reconstructed track is required to penetrate the 
first four BGO layers. The pre-selection efficiency is shown in Fig.~S1 of
the {\tt Supplemental Material}. We can see that the efficiency reaches about
80\% for energies higher than a few GeV. At low energies the efficiency drops
quickly, due mainly to the trigger efficiency.

After these pre-selections, the main background for CREs is protons. 
We use a particle identification (PID) parameter, defined as 
${\rm PID} = F(E)[\log(R_{r}) \cdot \sin\theta + \log(R_{l}) \cdot \cos\theta]$ 
to distinguish CREs from protons, where $F(E)$ is an energy decoupling 
polynomial, ${\theta}$ is a rotation angle, ${R_{r}}$ and ${R_{l}}$ are 
parameters describing the radial and longitudinal extensions of the shower 
in the calorimeter \cite{2021ApJ...920L..43A}. Fig.~S2 of the 
{\tt Supplemental Material} illustrates the distributions of the PID 
parameter for two energy bins, $3.17-3.44$ GeV and $7.35-8.00$ GeV. 
CREs would typically have ${\rm PID}\leq 2$. To properly estimate the 
proton background, we carry out a fit to the data with Monte Carlo (MC) 
simulation templates of CREs and protons. Since CMEs affect electrons 
and protons differently \cite{2018ApJ...853...76M}, the background is 
estimated independently for each energy and time bin. We find that the 
proton background is about 2\% to 8\% for energies from 2 to 20 GeV, with 
about $15\%$ relative changes (differ case by case) during the FD period.

The CRE flux is calculated as
\begin{equation}
    I_{i,j}=\frac{N_{i,j}}{{\Delta} E_{i} A_{i} T_{i,j} \eta_{i}},
    \label{eq2}
\end{equation}
where ${N_{i,j}}$ is the number of selected CREs after the background
subtraction in the ${i}$th energy bin and ${j}$th time bin, ${T_{i,j}}$ 
is the corresponding exposure time, ${\Delta E_{i}}$ is the energy bin 
width, ${A_{i}}$ is the geometric acceptance, and $\eta_i$ is the total 
efficiency. To avoid the shielding effect from the geomagnetic field, 
we require that the particle's rigidity is higher than 1.2 times 
of the VRC values. Since the VRC value varies with latitude, the rigidity 
threshold results in variations of the exposure time with time and energy 
(see Fig.~S3 of the {\tt Supplemental Material}). On average, 
the effective exposure time is approximately 80\% of the total time for 
energies above 15 GeV, and decreases gradually to about 30\% at 2 GeV.

\begin{figure*}
\centering
\includegraphics[width=1.0\textwidth]{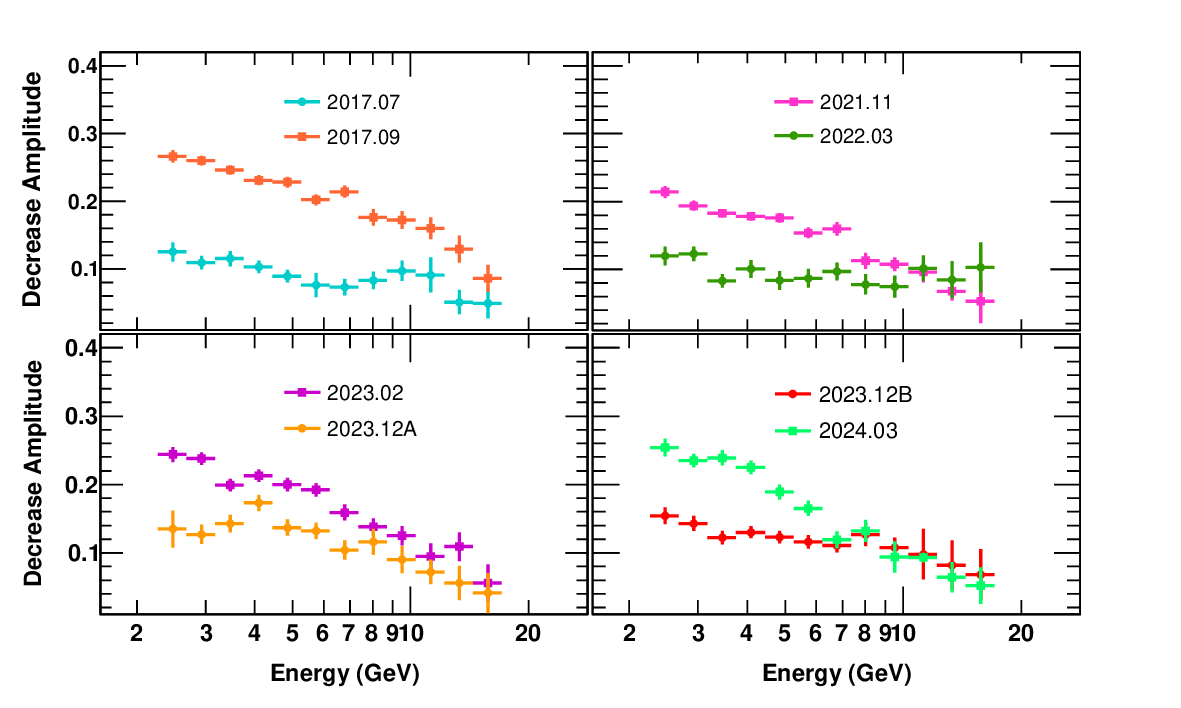}
\caption{The decrease amplitude as a function of energy, for the eight FDs
observed by DAMPE.}
\label{fig:amp}
\end{figure*}

\begin{figure*}
\centering
\includegraphics[width=1.0\textwidth]{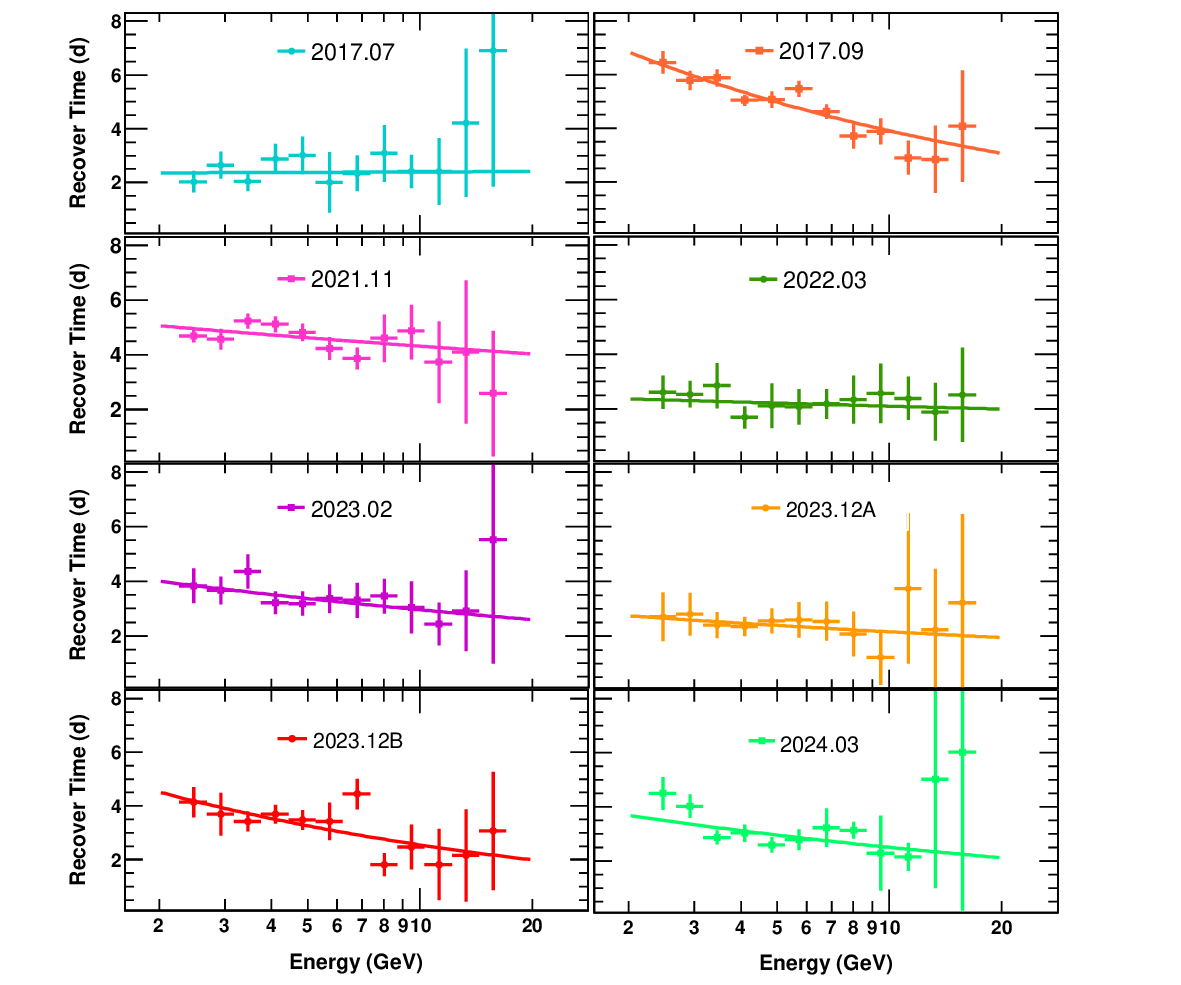}
\caption{The recover time as a function of energy, for the eight FD observed 
by DAMPE. The solid line in each panel shows the best-fit power-law result.}
\label{fig:tau}
\end{figure*}

Since we focus on the relative change of the CRE fluxes at a time scale of 
about one month, the main systematic uncertainties are the efficiency 
stabilities. The efficiency fluctuation is estimated to be $<0.15\%$ within 
one month after correcting the temperature effect \cite{2019APh...106...18A}. 
It has also been shown that there is a long-term evolution of the trigger 
efficiency due to radiation damage, electronics aging, and gain decrease 
of photomultipliers, which results in a decrease of the HET efficiency by 
about 5\% per year at 2 GeV and smaller with increasing energies 
\cite{2024NIMPA106969815L}. The systematic uncertainties are much smaller 
than the statistical uncertainties, which are about 3\% at 2 GeV and about 
7\% at 20 GeV.

{\it Results of FDs.} --- 
Fig.~\ref{fig:2021FD} shows the measured CRE fluxes with a time resolution 
of 6 hours in 4 energy bins for the FD event in November, 2021. Also shown 
are the NM data from the OULU station (red), and $T_0$ count rate of DAMPE 
(green) for the McIlwain L-values \cite{1994CASSS..10.....W} of $3.8-4.7$ 
which roughly corresponds to the VRC value of the OULU NM station. 
The fluxes 20 days prior to the FD are used as reference for normalizations. 
Note that, although the cutoff rigidities of the OULU station and the DAMPE 
$T_0$ events are similar, they represent essentially different particle 
energies since NMs record secondary neutrons produced by cosmic rays with 
much higher energies. This explains the differences of their decrease 
amplitudes.

As can be found from Fig.~\ref{fig:2021FD}, the CRE fluxes experience a fast 
decrease at UT 21:00 on November 3, 2021, reach the minimum at UT 21:00 on 
November 4, and gradually recover to the normal level in about 2 weeks. 
The $T_0$ rate follows a trend similar to that of CRE fluxes. Different from 
the September 2017 event \cite{2021ApJ...920L..43A}, no precursor was observed 
for in this event. The NM profile reaches its minimum at similar time as the 
$T_0$ count rate, but is about 3 hours later than the minimum time of the CRE 
fluxes. The NM profile also shows a faster recovery (with characteristic 
timescale of $\sim2$ days) than the CRE fluxes and the $T_0$ count rate 
(with characteristic timescale of $\sim 4.5$ days).
Both the NM and the DAMPE $T_0$ data experience two-step decrease features,
as marked by arrows. Possible explanations of the two-step features could be 
found in Ref.~\cite{2004JGRA..109.2117I}. We note that the $T_0$ rate exhibits 
a larger decrease amplitude in the first step and a smaller amplitude in the 
second step, in contrary to those of NM data. Such differences may again be 
due to different energies. The time profiles for other events can be found 
in Fig.~S6 of the {\tt Supplemental Material}.

The recovery phase of the FD is important in revealing the properties of the
interplanetary environment disturbed by the CME. The following function is 
employed to fit the recovery profile of an FD: 
\begin{equation}
\frac{I_{t}}{I_{0}}=1-A_{e} \exp\left(-\frac{t-t_{m}}{\tau}\right),
    \label{eq3}
\end{equation}
where ${I_{t}}$ is the measured CRE flux at time $t$, ${I_{0}}$ is the 
normalization flux before the FD, ${A_{e}}$ is the decrease amplitude, 
${t_{m}}$ is the time of minimum, and ${\tau}$ is the characteristic recovery 
time.

\begin{figure}
\centering
\includegraphics[width=1.0\linewidth]{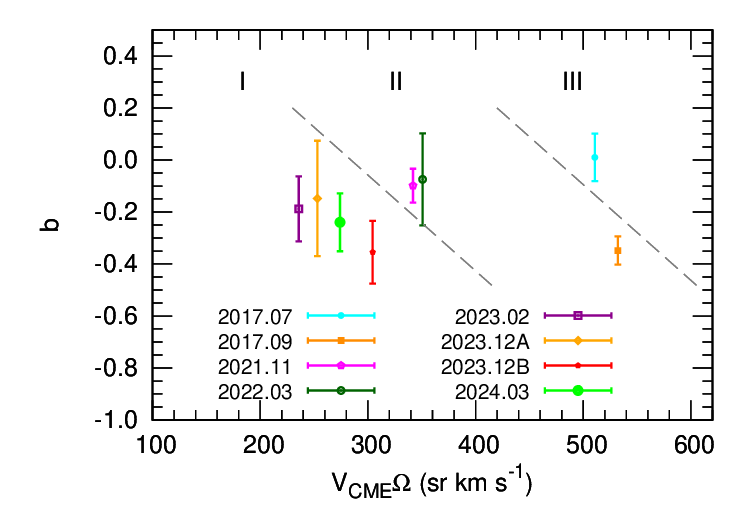}
\caption{Possible relationship between $b$ and $V_{\rm CME}\cdot\Omega$ as 
revealed by the DAMPE detected FDs. The dashed lines divide the CMEs 
schematically into head-on events (I), glacing events (III), and intermediate
ones (II).}
\label{fig:tau_VOmega}
\end{figure}

The derived decrease amplitudes and recovery times, for each energy
bin, are shown in Fig.~\ref{fig:amp} and Fig.~\ref{fig:tau}. 
Note that when performing the fitting, the systematic uncertainties
are conservatively included via a quadratic sum with the statistical
uncertainties. The decrease amplitude becomes smaller with the increasing 
energy. This can be explained by the fact that particles with 
higher energy are less affected by the magnetic turbulence.
The energy dependence of the recovery time, however, shows diverse 
properties. For some events, the recovery time becomes shorter with the 
increase of energy (e.g., the 2017-09 event), while for others it remains 
nearly constant for all energies (e.g., the 2017-07 event). These diverse 
behaviors may be related with properties of CMEs causing the FDs. We further 
fit the energy dependence of the recovery time with a power-law function,
$\tau=aE^b$, and derive the power-law index $b$ for each event. The results 
of the parameter $b$ for the 8 events are shown in Fig.~\ref{fig:tau_VOmega}. 

To investigate the behaviors of the recovery time, we also show in 
Fig.~\ref{fig:tau_VOmega} the possible relation between $b$ and a combined 
parameter, $V_{\rm CME}\cdot\Omega$, for these events. The CME velocity 
$V_{\rm CME}$ reflects the strength of the disturbance, and $\Omega$ is the 
solid angle of the CME which reflects the spatial size affected by the CME.
The CME parameters are extracted from the Wang-Sheeley-Arge (WSA)-Enlil solar 
wind prediction 
model\footnote{https://www.swpc.noaa.gov/products/wsa-enlil-solar-wind-prediction}
\cite{1990ApJ...355..726W,2003AIPC..679..190A,2013SpWea..11...57M}.
The model reconstructs the CME speed ($V_{\text{CME}}$) and the solid angle 
($\Omega$) in a region of 5 to 20 solar radii using corona graphs from three 
different satellites, including STEREO-B COR2, STEREO-A COR2, and LASCO C2. 
The projection effect has been taken into account. Another potentially 
important parameter affecting the FDs of GCRs is the ejection direction of 
the CME \cite{2016ApJ...827...13Z}. We classify the CMEs into
head-on events (I), glancing events (III), and intermediate events (II) 
(see Fig.~S5 of the {\tt Supplemental Material} for a schematic plot of the 
classification). Note that the boundaries among these categories are not 
sharply defined, and the current classification is just for illustrative 
purpose. As can be seen in Fig.~\ref{fig:tau_VOmega}, for classes I and II 
events, anti-correlations between the slope $b$ and $V_{\rm CME}\cdot\Omega$ 
is observed. Only one class III FD event has been detected at present, and the 
relationship between $b$ and $V_{\rm CME}\cdot\Omega$ is not clear yet. 
An exception is the March 2022 event, which should be classified as a head-on 
event based on the CME parameters (see Table~S1 of the {\tt Supplemental 
Material}), but its ($b$, $V_{\rm CME}\cdot\Omega$) values place it in region
II as labelled in Fig.~\ref{fig:tau_VOmega}. The precise classification of this 
event is ambiguous since the decrease amplitude is small, leading to a 
relatively large uncertainty of the slope parameter $b$. Roughly, stronger CMEs
(faster in speed and wider in spread) would result in stronger energy 
dependence of the recovery time.

The transport of a charged particle in the heliosphere can be described by 
the Parker's equation \cite{1965P&SS...13....9P} (see the {\tt Supplemental 
Material}). CMEs and other solar activities may disturb the interplanetary 
space, leading to perturbation of the propagation of GCRs. This process can 
be modeled by a fast-moving diffusion barrier \cite{2018ApJ...860..160L}. 
More details about the setting of the diffusion barrier can be found in 
the {\tt Supplemental Material}. We adopt the stochastic differential 
equation (SDE) method to solve Parker's equation with the disturbance by 
CMEs \cite{1999ApJ...513..409Z,2011ApJ...735...83S}. In the modeling, the 
basic parameters of the CMEs, such as the velocity, angular extension, and 
direction, are from the WSA-ENLIL model. The other parameter is the diffusion
coefficient inside the barrier, which is adjusted to fit the measurements. 
As shown in Fig.~S6 of the {\tt Supplemental Material}, the time 
profiles of the CRE fluxes around the FDs can be reproduced by this model. 
The comparison of the energy-dependence slope parameter $b$ of the recovery 
time between DAMPE observations and simulations is given in Fig.~S7 
of the {\tt Supplemental Material}, which also shows a good consistency. 
The simulation indicates that the recovery behaviors of FDs depend on multiple
factors, including the motion and geometry of the disturbed regions. Thus the
diverse properties of the CRE FDs observed by DAMPE reflect the complicated 
propagation processes of particles in the interplanetary space. Note further
that, the FD events reported in this analysis occurred between 2017 and 2024,
during which the solar magnetic polarity was always 
positive\footnote{A positive solar magnetic polarity corresponds to an 
outward-pointing heliospheric magnetic field in the Sun's northern hemisphere
and an inward-pointing magnetic field in the southern hemisphere.} ($A>0$). 
We have a consistent drift direction configuration for all events, 
isolating the additional complexity due to the polarity reversal.

{\it Conclusion.} --- 
In this work we study the FDs of CREs recorded by DAMPE, from January, 2016 
to March, 2024. Different from most of existing studies which focus
on the nuclear component of GCR FDs, this is the first systematic investigation
of FDs of the electron and positron component. In the total 8 FD events
detected, the largest decrease amplitude reaches about 30\%. The energy 
dependence of the decrease amplitude and recovery time, from 2 GeV to 20 GeV, 
is well measured. We find that the energy dependence of the recovery time show 
diverse behaviors, which span from strongly energy-dependent to nearly 
energy-independent. Our results further suggest that this diversity is likely
due to a combined effect of the CME velocity, angular extension, and 
propagation direction. Depending on the ejection direction of the CME 
relative to the Earth, an anti-correlation between the energy dependence 
slope of the recovery time and the velocity times solid angle of the CME 
is revealed by the data. Utilizing a diffusion barrier model, we can 
account for the measurements under the framework of Parker's transport 
equation, based on observed parameters of the CMEs.

With the sunspot number approaching its maximum during the solar cycle 25, 
more FD events are expected to be detected, which will be very helpful in 
testing the phenomenae found in this study. Particularly, additional events 
belonging to with classes II and III will be especially important. 
Furthermore, the comparative studies of FDs of CREs and protons or other 
nuclei, as in Ref.~\cite{2018ApJ...853...76M} but with a significantly 
enlarged sample are of great interest and will be carried out in future works.
Finally, the joint analysis of the FD events and the sun shadow by groundbased
cosmic ray experiments \cite{2024Innov...500695} will be crucial in improving 
our understanding of impacts of solar activities on the interplanetary 
magnetic environment and the particle propagation in the heliosphere.

The data that support the findings of this article are openly 
available\footnote{https://doi.org/10.57760/sciencedb.space.02998}.

\section*{Acknowledgments}
The DAMPE mission was funded by the strategic priority science and technology 
projects in space science of Chinese Academy of Sciences (CAS). 
In China, the data analysis is supported by the National Key Research and 
Development Program of China (No. 2022YFF0503302), the National Natural Science
Foundation of China (Nos. 12220101003 and 12321003), the CAS Project for Young 
Scientists in Basic Research (No. YSBR-061), and the Youth Innovation Promotion 
Association of CAS. 
In Europe, the activities and data analysis are supported by the Swiss National 
Science Foundation (SNSF), Switzerland, the National Institute for Nuclear 
Physics (INFN), Italy, and the European Research Council (ERC) under the 
European Union's Horizon 2020 research and innovation programme (No. 851103).

\bibliography{refs}
\bibliographystyle{apsrev}

\clearpage



\onecolumngrid

\setcounter{figure}{0}
\renewcommand\thefigure{S\arabic{figure}}
\setcounter{table}{0}
\renewcommand\thetable{S\arabic{table}}





\begin{center}
{\Large Supplemental Material of ``Diverse properties of electron Forbush 
decreases revealed by the Dark Matter Particle Explorer''}\\
(The DAMPE collaboration)\\
\end{center}

\section{Properties of 8 CRE FDs and the associated CMEs}
Table \ref{tab:8FD} gives the main parameters of the 8 FDs discussed in this 
work, including the DAMPE measured CRE fluxes, the CME parameters from the 
WSA-ENLIL model, and the parameters of the diffusion barrier used in our 
simulation.

\begin{table*}[!htbp]
\centering
\caption{Main parameters of the 8 FDs. From left to right, they are, the time 
at which the DAMPE $T_0$ rate reaches the minimum (Time), the maximum decrease 
amplitude of DAMPE CRE fluxes ($A_e^{\rm max}$), the recover time of DAMPE CRE 
at 2.3 GeV ($\tau$), the slope of energy dependence of recover time ($b$), the 
CME speed ($V_{\rm CME}$), the half angle width of CME ($\Theta$), the 
reduction factor of the diffusion coefficient in the barrier ($\rho$), 
the location of the Earth in the CME coordinate ($\theta_{0}$,$\phi_{0}$), 
the latitude and longtitude angular widths of CME ($\theta_{\rm br}$,
$\phi_{\rm br}$). The symbol in parenthesis after $\phi_0$ gives the 
classification of the event as defined in this work.}
\resizebox{0.8\textwidth}{22mm}{
\begin{tabular}{c|c|c|c|c|c|c|c|c|c|c}
\hline
\multicolumn{4}{c|}{DAMPE CRE Observations}&\multicolumn{2}{c}{CME Parameters}&\multicolumn{5}{|c}{Simulation Parameters} \\
\hline
\makebox[0.14\textwidth][c]{Time (UTC)} & \makebox[0.1\textwidth][c]{$A_{e}^{\rm max}$} & \makebox[0.1\textwidth][c]{$\tau$ (day)} & \makebox[0.1\textwidth][c]{$b$} & \makebox[0.12\textwidth][c]{$V_{\rm CME}$ (km s$^{-1}$)} & \makebox[0.05\textwidth][c]{$\Theta$($^{\circ}$) } & \makebox[0.04\textwidth][c]{$\rho$} & \makebox[0.04\textwidth][c]{$\theta_0$($^{\circ}$)} & \makebox[0.04\textwidth][c]{$\phi_0$($^{\circ}$)} & \makebox[0.04\textwidth][c]{$\theta_{\rm br}$($^{\circ}$)} & \makebox[0.04\textwidth][c]{$\phi_{\rm br}$($^{\circ}$)} \\
\hline
2017-07-17 04:30 & $0.125\pm0.014$ & $2.03\pm0.39$ & ~$0.010\pm0.091$ & 1199 & 57 & 4.25 & 70 & $-50$ (III) & 100 & 120 \\
2017-09-08 13:30 & $0.266\pm0.009$ & $6.46\pm0.42$ & $-0.348\pm0.054$ & 1611 & 48 & 8.25 &90 & 45 (II)  & 90  & 120 \\
2021-11-04 21:30 & $0.214\pm0.009$ & $4.69\pm0.24$ & $-0.099\pm0.065$ & 1397 & 41 & 7.00 &90 & 30 (II)  & 90  & 90 \\
2022-03-14 19:30 & $0.123\pm0.014$ & $2.61\pm0.61$ & $-0.074\pm0.018$ & 983  & 50 & 6.50 &100 & 0 (I)  & 100  & 100 \\
2023-02-28 14:30 & $0.244\pm0.011$ & $3.83\pm0.64$ & $-0.187\pm0.125$ & 1237 & 36 & 5.00 &90 & 0 (I)  & 60  & 80 \\
2023-12-01 18:30 & $0.137\pm0.027$ & $2.72\pm0.89$ & $-0.148\pm0.222$ & 944  & 43 & 5.50 &90 & 0 (I)  & 90  & 90 \\
2023-12-17 22:30 & $0.154\pm0.012$ & $4.15\pm0.56$ & $-0.355\pm0.121$ & 921  & 48 & 7.25 &90 & 10 (I)  & 100  & 90 \\
2024-03-24 22:30 & $0.254\pm0.013$ & $4.49\pm0.60$ & $-0.239\pm0.111$ & 1438 & 36 & 10.0 &100 & 0 (I)  & 90  & 80 \\
\hline
\end{tabular}
}
\label{tab:8FD}
\end{table*}

\section{Pre-selection efficiency}
\begin{figure*}[!ht]
\centering
\includegraphics[width=0.6\linewidth]{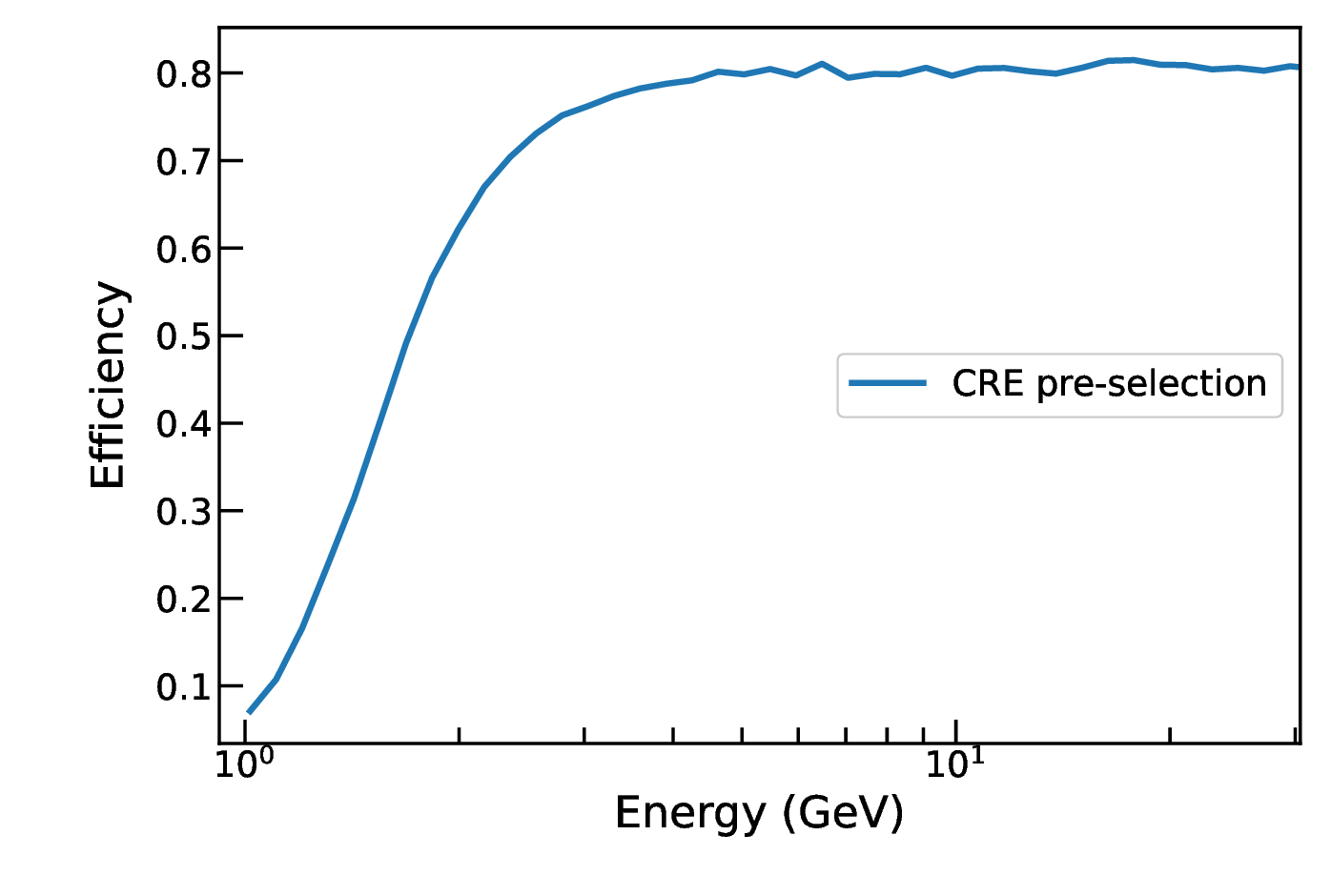}
\caption{The CRE pre-selection effciency as a function of reconstructed energy.}
\label{fig:eff}
\end{figure*}

The pre-selections include the trigger selection, the charge selection, 
the track selection, and the shower containment requirement. The selection 
efficiency as a function of energy is shown in Fig.~\ref{fig:eff}.

\section{Background subtraction}
\begin{figure*}[!htb]
\centering
\includegraphics[width=\textwidth]{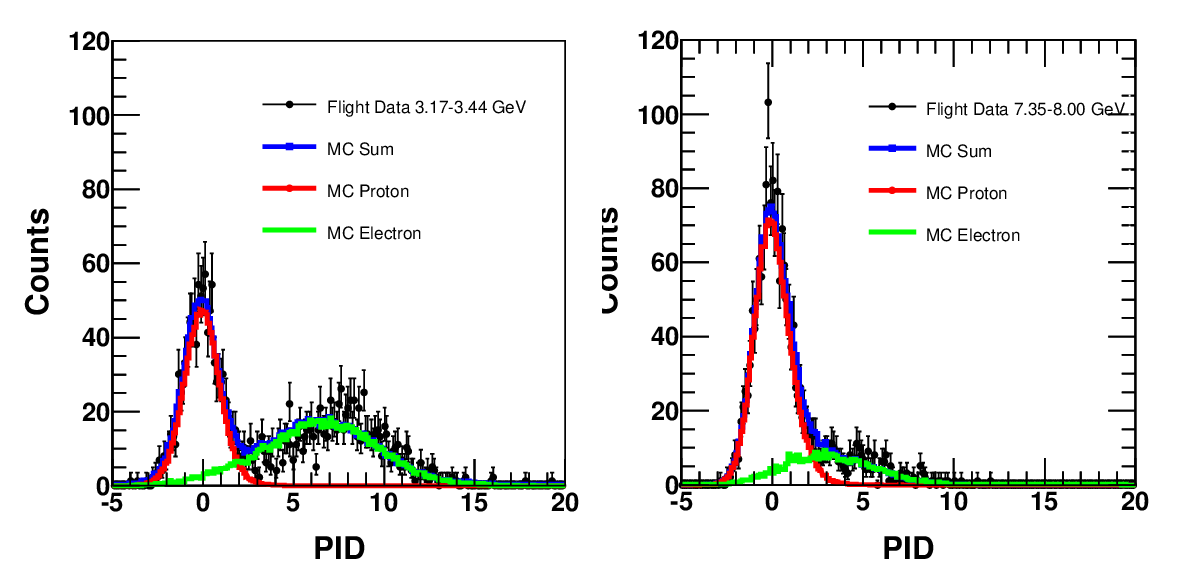}
\caption{The PID distributions of the flight data (black dots) and MC 
simulations (red for electrons, green for protons, and blue for their sum). 
The left panel is for the energy bin of $3.17 - 3.44$ GeV, and the right 
panel is for the energy bin of $7.35-8.00$ GeV.}
\label{fig:PID}
\end{figure*}

After pre-selections, the data sample includes mainly CREs and protons. 
To distinguish CREs from the proton background, a particle identification 
(PID) variable is defined \cite{2021ApJ...920L..43A} to describe the shower 
development in the calorimeter. Fig.~\ref{fig:PID} shows the PID distributions 
in two energy bins. Curves in red and green are MC simulated PID distributions 
of CREs and protons, and the blue ones are the sum of these two components, 
obtained by a fit to the data using the MC templates. The proton contamination 
is then obtained and subtracted from the data sample. 

\section{Exposure correction at low energies}
\begin{figure*}[!htb]
\centering
\includegraphics[width=0.7\textwidth]{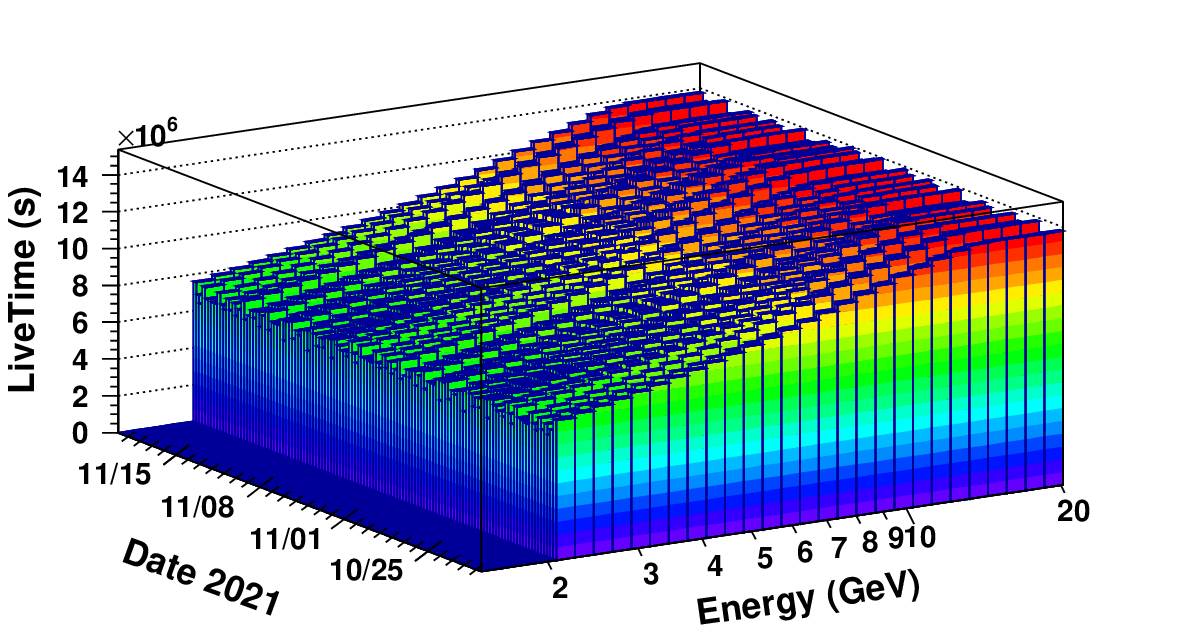}
\caption{Exposure time of DAMPE as a function of time and energy.}
\label{fig:exp}
\end{figure*}

Due to the geomagnetic field, low energy CRs will get shielded and can not 
reach the orbit of DAMPE. This corresponds to a cutoff of the low energy 
spectrum of particles. To avoid the impact of the cutoff effect, a threshold 
of the particle rigidity of 1.2 times of the VRC value is set, which results 
in latitude-dependent (and consequently time-dependent) and energy-dependent 
exposure time. For each energy and time interval, the exposure time is 
calculated as the sum of the satellite travel time when this given rigidity 
satisfies the above threshold condition. The time when the satellite passes 
through the South Atlantic Anomaly (SAA) region, when the instrument switches 
to the calibration mode, and the dead time of the data acquisation system, 
is also excluded. Fig.~\ref{fig:exp} shows an illustration of the exposure time 
from 2 GeV to 20 GeV for the time from November 1 to November 31, 2021. We can
see that the exposure time is almost constant above 12 GeV, and decreases at
lower energies due to the geomagnetic cutoff effect. 

\section{Electron transport in the heliosphere}

The transport of particles in the heliosphere is described by the Parker's 
equation \cite{1965P&SS...13....9P}:
\begin{equation}
    \frac{\partial f}{\partial t} = -({\vec{V}}_{\rm sw} 
    + \langle{\vec{V}}_{\rm d}\rangle)\cdot\bigtriangledown{f} 
    + \bigtriangledown\cdot(K^{(\rm s)}\cdot\bigtriangledown{f}) 
    + \frac{1}{3}(\bigtriangledown\cdot{\vec{V}}_{\rm sw})
      \frac{\partial{f}}{\partial{\ln p}},
    \label{eq5}
\end{equation}
where $f(\vec{r},p,t)$ is the particle distribution function, $p$ is momentum, 
${\vec{V}_{\rm sw}}$ is the latitude-dependent solar wind speed, 
$\langle{\vec{V}}_{\rm d}\rangle$ is the pitch-angle averaged drift velocity 
which has the following form \cite{1977ApJ...213..861J,1989ApJ...339..501B} 
\begin{equation}
    \langle\vec{V}_{\rm d}\rangle = \frac{pv}{3q}\bigtriangledown\times\frac{\vec{B}}{B^{2}},
    \label{appeq2}
\end{equation}
where $q$ and $v$ are charge and velocity of GCR particles, and $\vec{B}$ 
is the interplanetary magnetic field (IMF). The variable $K^{(\rm s)}$ is the
symmetric diffusion tensor, aligned in the IMF coordinate, whose diagonal 
components are the parallel diffusion coefficient $K_{||}$, and the 
perpendicular diffusion coefficiens $K_{\bot\theta}$ and $K_{\bot{r}}$.

\begin{figure*}[!htb]
\centering
\includegraphics[width=0.9\textwidth]{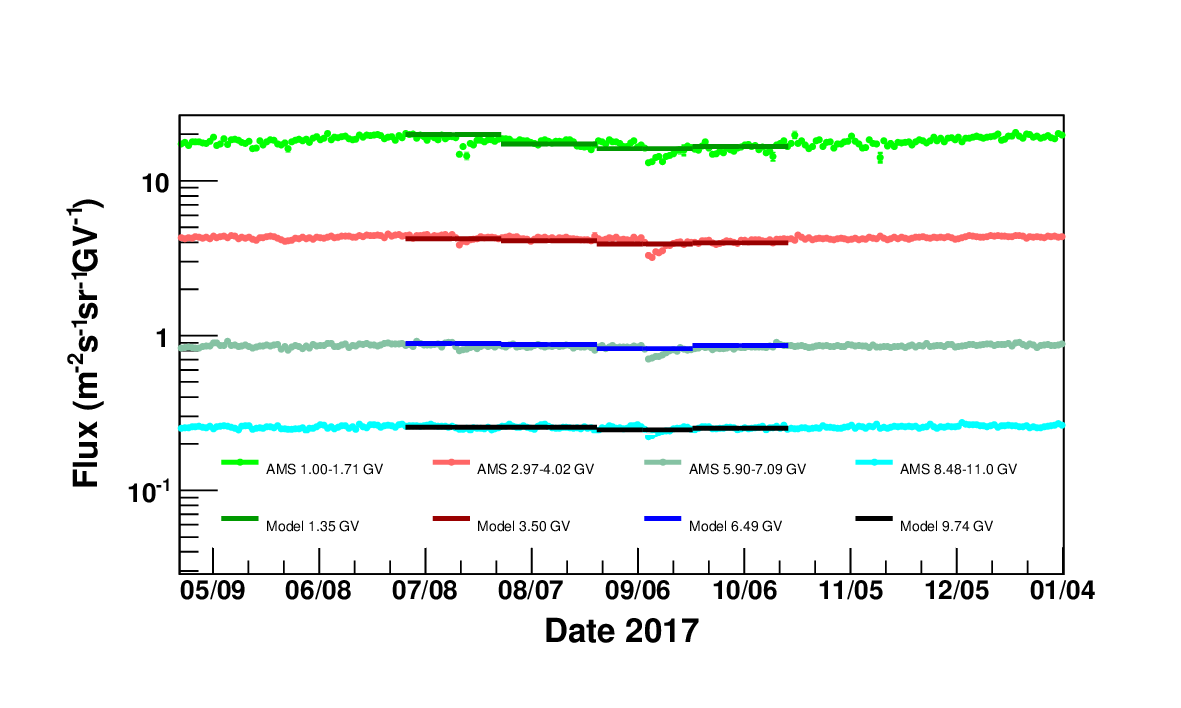}
\caption{Simulation of the solar modulation effect on electrons, compared 
with the AMS-02 daily electron fluxes \cite{2023PhRvL.130p1001A}, for Bartel 
rotations from 2509 to 2512 in 2017.}
\label{fig:SM}
\end{figure*}

\begin{figure*}[!htb]
\centering
\includegraphics[width=0.55\linewidth]{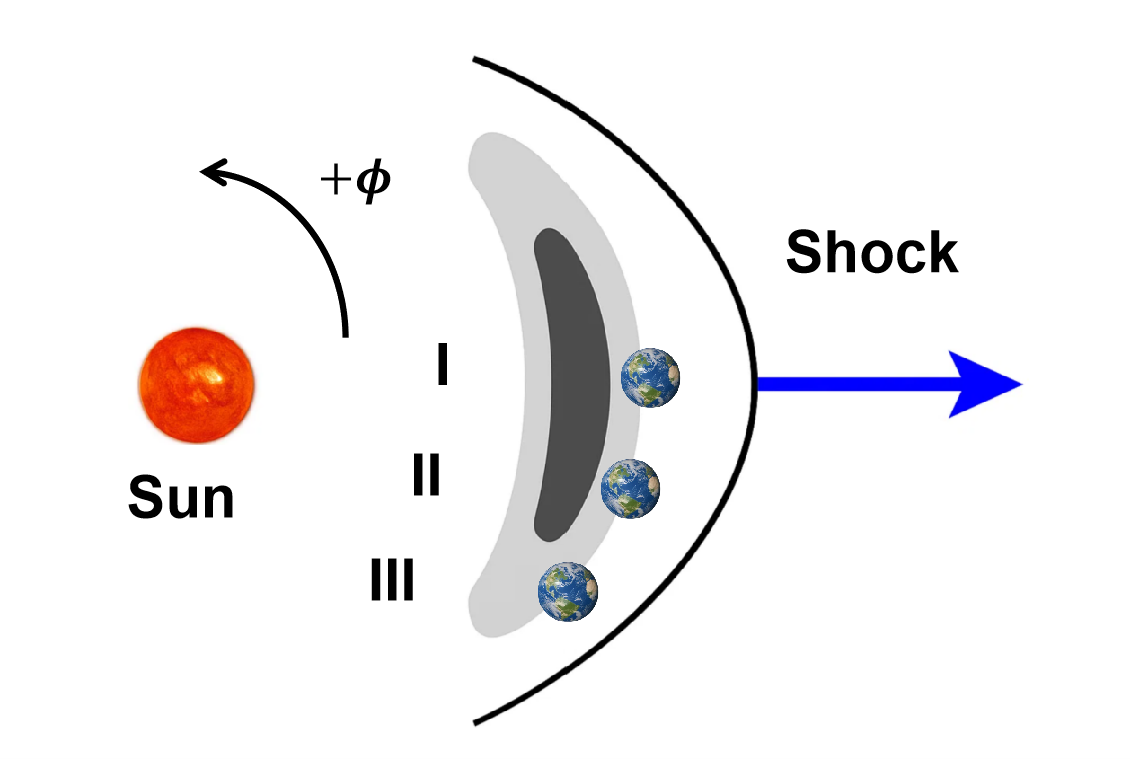}
\caption{Schematic plot to show the relative location of the Earth in the CME
for the three types of events.}
\label{fig:class}
\end{figure*}

In this work the diffusion coefficients are parameterized as \cite{2019ApJ...878....6L})
\begin{equation}
  K_{i}=K_{i0}\beta\frac{B_{0}}{B}\left(\frac{P}{P_{k}}\right)^{a}
  \left[1+\left(\frac{P}{P_{k}}\right)^{\frac{b-a}{c}}\right]^c,
  \label{appeq3}
\end{equation}
where $i$ represents $||$, $\bot\theta$, or $\bot r$; $\beta$ is the particle 
velocity in speed of light, $B$ is the strength of the IMF, $B_{0} \equiv 1$ nT,
$K_{i0}$ is the normalization diffusion coefficient varying with time, $p_{k}$
is the break momentum of the diffusion coefficient, $a$, $b$ are the power-law
indices before and after the break, and $c$ characterizes the smoothness of the 
break.

The IMF is described by the standard Parker field as
\begin{equation}
    \vec{B}(r,\theta) = \frac{AB_{\oplus}}{r^{2}}\left(\hat{e}_{r}-\frac{r\Omega{\sin\theta}}{V_{\rm sw}}\hat{e}_{\theta}\right) [1-2H(\theta-\theta_{\rm cs})],
    \label{addeq5}
\end{equation}
where $B_{\oplus}$ is an average of 13 months magnetic field strength before 
the specific time at 1 AU, $A$ describes the polarity of the IMF, defined as 
$A=\pm{1}/\sqrt{1+(\Omega/V_{\rm sw})^{2}}$, $\theta_{\rm cs}$ determines the 
polar extent of the Heliospheric Current Sheet (HCS), and $\Omega$ is the 
angular velocity of the sun. 

The Parker's equation is solved numerically with the stochastic differential
equation (SDE) technique \cite{1999ApJ...513..409Z,2011ApJ...735...83S}. The 
parameters of the diffusion coefficient are derived through fitting the 
electron fluxes measured by AMS-02 \cite{2023PhRvL.130p1001A}, using the local
interstellar spectrum of electrons derived in Ref.~\cite{2021APh...12402495Z} 
(see Fig.~\ref{fig:SM} for an example of the model fitting results in 2017). 
The relatively small portion of positrons has been omitted in the calculation.

\section{The diffusion barrier model of FD}
\begin{figure*}[!htb]
\centering
\includegraphics[width=0.9\textwidth]{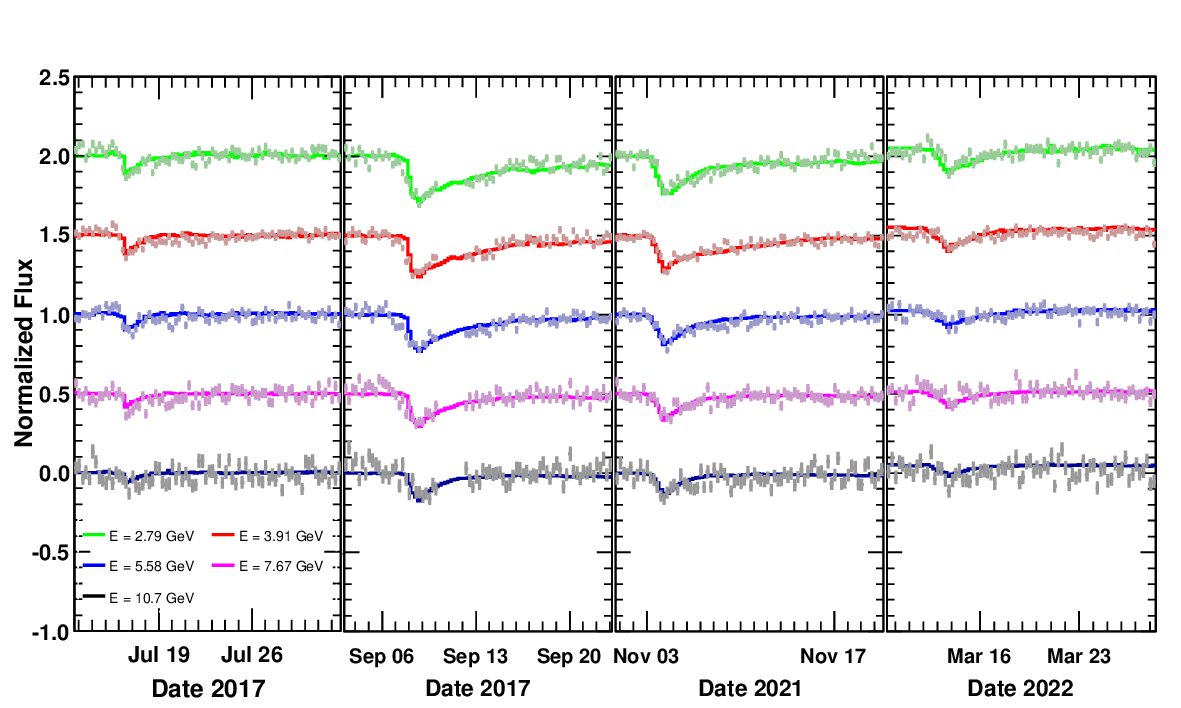}
\includegraphics[width=0.9\textwidth]{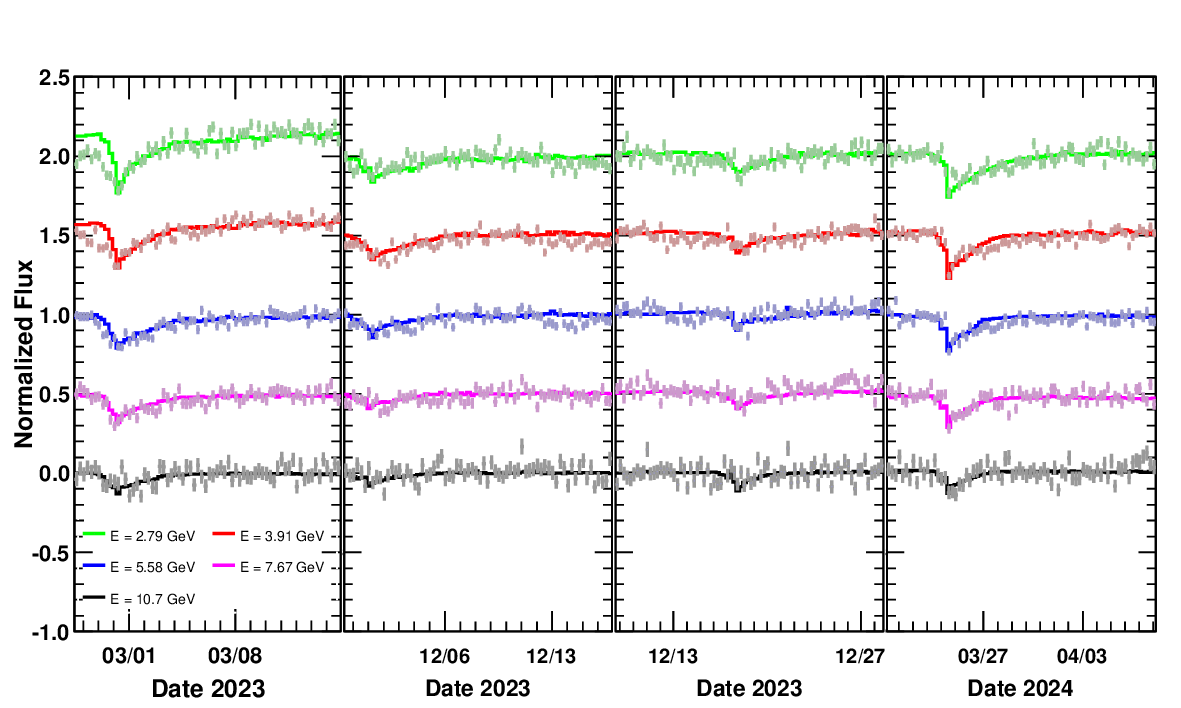}
\caption{Comparison of the DAMPE observed CRE time profiles (dots) with the 
simulation results (lines) for 5 selected energy bins.}
\label{fig:simu}
\end{figure*}

\begin{figure*}[!htb]
\centering
\includegraphics[width=0.7\linewidth]{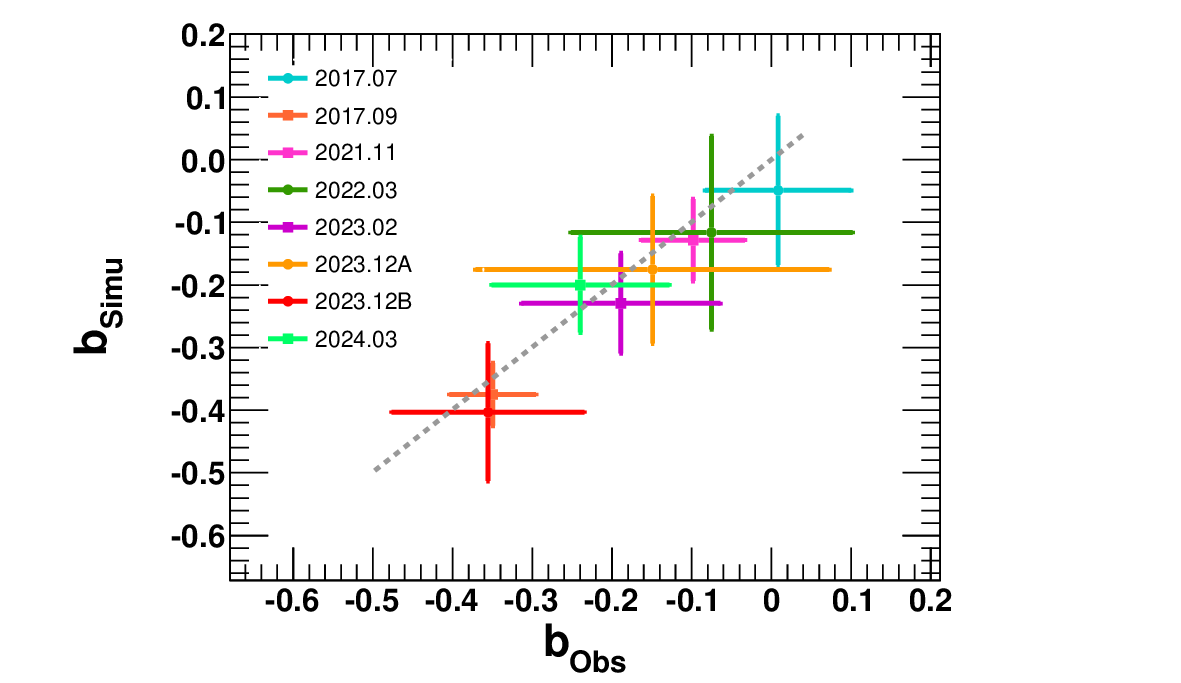}
\caption{The slope parameter $b$ derived from DAMPE observations 
($b_{\rm Obs}$) and simulations ($b_{\rm Simu}$).}
\label{fig:b}
\end{figure*}

The CME is described by a three dimensional diffusion barrier, in which the
diffusion coefficient is smaller than the interplanetary value
\cite{2018ApJ...860..160L}. The diffusion coefficient inside the barrier is 
written as
\begin{equation}
    K{'}_{i}=\frac{K_{i}}{1+\rho{h(\theta)f(r)g(\phi)}},
\end{equation}
where $\rho$ is the reduction factor, $h(\theta), g(\phi)$, and 
$f(r)$ describes the geometry of the diffusion barrier, as
\begin{equation}
    h(\theta)=e^{-\left(\frac{\theta-\theta_{0}}{\theta_{\rm br}}\right)^{10}},
\end{equation}
\begin{equation}
    g(\phi)=e^{-\left(\frac{\phi-\phi_{0}}{\phi_{\rm br}}\right)^{10}},
\end{equation}
\begin{equation}
 f(r)=\left\{
 \begin{aligned}
 1-\frac{r-r_{\rm cen}}{r_{a}}&,&r_{\rm cen}<r<r_{\rm fr} \\
 \frac{r-r_{\rm end}}{r_b}&,& r_{\rm end}<r\leq{r_{\rm cen}}\\
 0& ,& {\rm others}
 \end{aligned}
 \right.,
\end{equation}
where $(\theta_0,\phi_0)$ are the direction of the CME center. 
For $\theta_0=\pi/2$ and $\phi_0=0$, the CME hits the Earth head on. 
Parameter $\theta_{\rm br}$ and $\phi_{\rm br}$ describes the extension 
of the barrier. The radical location of the front, center, and end of the 
CME are depicted as $r_{\rm fr}$, $r_{\rm cen}$, and $r_{\rm end}$, and 
the widths of the leading and trailing parts of the barrier are 
$r_{a}=r_{\rm fr}-r_{\rm cen}$ and $r_{b}=r_{\rm cen}-r_{\rm end}$. 
The same SDE method is employed to calculate the time-dependent electron 
fluxes including the diffusion barrier. The parameters used to fit the 
DAMPE data are given in Table \ref{tab:8FD}. A comparison between the 
observed time profiles and simulated ones for the 8 FDs is shown in 
Fig.~\ref{fig:simu}. We also fit the slopes of the energy-dependence of 
the recovery time for the simulation results, which are shown in 
Fig.~\ref{fig:b} with a comparison with observations. Here the 
uncertainties of $b_{\rm Simu}$ are obtained through scaling down
the simulation sample to similar event statistics with the flight 
data, which makes the comparison between the observations and 
simulations more direct.

\end{document}